\documentclass[twocolumn]{aastex631}

\usepackage[normalem]{ulem}
\usepackage{multirow}

\def\Lsun{L_\odot}
\def\Msun{M_\odot}
\def\MJ{M_\mathrm{J}}
\def\Mp{M_\mathrm{p}}

\def\Mstar{M_{*}}
\def\Macc{\dot{M}_\mathrm{acc}}
\def\RJ{R_\mathrm{J}}
\def\Rp{R_\mathrm{p}}
\def\Teff{T_{\text{eff}}}

\def\Lacc{L_\mathrm{acc}}
\def\LHa{L_\mathrm{H\alpha}}
\def\Lline{L_\mathrm{line}}

\def\Av{A_\mathrm{V}}

\begin{document}

\title{JWST/NIRC\lowercase{am} Imaging of Young Stellar Objects. II. Deep Constraints on Giant Planets and a Planet Candidate Outside of the Spiral Disk Around SAO~206462 
}

\correspondingauthor{Gabriele Cugno}
\email{gcugno@umich.edu}

\author[0000-0001-7255-3251]{Gabriele Cugno}
\affiliation{Department of Astronomy, University of Michigan, Ann Arbor, MI 48109, USA}

\author[0000-0002-0834-6140]{Jarron Leisenring}
\affiliation{Department of Astronomy and Steward Observatory, University of Arizona, USA}

\author[0000-0002-4309-6343]{Kevin R. Wagner}
\affiliation{Department of Astronomy and Steward Observatory, University of Arizona, USA}

\author[0009-0007-3210-4356]{Camryn Mullin}
\affiliation{Department of Physics and Astronomy, University of Victoria, Victoria, BC, V8P 5C2, Canada}

\author[0000-0001-9290-7846]{Roubing Dong}
\affiliation{Department of Physics and Astronomy, University of Victoria, Victoria, BC, V8P 5C2, Canada}

\author[0000-0002-8963-8056]{Thomas Greene}
\affiliation{NASA Ames Research Center, MS 245-6, Moffett Field, CA 94035, USA}

\author[0000-0002-6773-459X]{Doug Johnstone}
\affiliation{NRC Herzberg Astronomy and Astrophysics, 5071 West Saanich Road, Victoria, BC, V9E 2E7, Canada}
\affiliation{Department of Physics and Astronomy, University of Victoria, Victoria, BC, V8P 5C2, Canada}

\author[0000-0003-1227-3084]{Michael R. Meyer}
\affiliation{Department of Astronomy, University of Michigan, Ann Arbor, MI 48109, USA}

\author[0000-0002-9977-8255]{Schuyler G. Wolff}
\affiliation{Department of Astronomy and Steward Observatory, University of Arizona, USA}

\author[0000-0002-5627-5471]{Charles Beichman}
\affiliation{Jet Propulsion Laboratory, Pasadena, CA, USA}

\author{Martha Boyer}
\affiliation{Space Telescope Science Institute, 3700 San Martin Drive, Baltimore, MD, 21218, USA}


\author[0000-0001-9886-6934]{Scott Horner}
\affiliation{NASA Ames Research Center, MS 245-6, Moffett Field, CA 94035, USA}

\author{Klaus Hodapp}
\affiliation{University of Hawaii, Hilo, HI 96720, USA}

\author{Doug Kelly}
\affiliation{Department of Astronomy and Steward Observatory, University of Arizona, USA}

\author{Don McCarthy}
\affiliation{Department of Astronomy and Steward Observatory, University of Arizona, USA}

\author[0000-0002-6730-5410]{Thomas Roellig}
\affiliation{NASA Ames Research Center, MS 245-6, Moffett Field, CA 94035, USA}

\author{George Rieke}
\affiliation{Department of Astronomy and Steward Observatory, University of Arizona, USA}

\author[0000-0002-7893-6170]{Marcia Rieke}
\affiliation{Department of Astronomy and Steward Observatory, University of Arizona, USA}

\author{John Stansberry}
\affiliation{Space Telescope Science Institute, 3700 San Martin Drive, Baltimore, MD, 21218, USA}

\author[0000-0002-6395-4296]{Erick Young}
\affiliation{Universities Space Research Association, 425 3rd St. SW, Suite 950, Washington, DC 20024, USA}

\begin{abstract}

We present JWST/NIRCam F187N, F200W, F405N and F410M direct imaging data of the disk surrounding SAO~206462. Previous images show a very structured disk, with a pair of spiral arms thought to be launched by one or more external perturbers. The spiral features are visible in three of the four filters, with the non-detection in F410M due to the large detector saturation radius. We detect with a signal-to-noise ratio of 4.4 a companion candidate (CC1) that, if on a coplanar circular orbit, would orbit SAO~206462 at a separation of $\sim300$ au, $2.25\sigma$ away from the predicted separation for the driver of the eastern spiral. According to the BEX models, CC1 has a mass of $M_\mathrm{CC1}=0.8\pm0.3~\MJ$. No other companion candidates were detected. At the location predicted by simulations of both spirals generated by a single massive companion, the NIRCam data exclude objects more massive than $\sim2.2~\MJ$ assuming the BEX evolutionary models. In terms of temperatures, the data are sensitive to objects with $\Teff\sim650-850$ K, when assuming planets emit like blackbodies ($\Rp$ between 1 and $3 \RJ$). From these results, we conclude that if the spirals are driven by gas giants, these must be either cold or embedded in circumplanetary material. In addition, the NIRCam data provide tight constraints on ongoing accretion processes. In the low extinction scenario we are sensitive to mass accretion rates of the order $\dot{M}\sim10^{-9} \MJ$ yr$^{-1}$. Thanks to the longer wavelengths used to search for emission lines, we reach unprecedented sensitivities to processes with $\dot{M}\sim10^{-7} \MJ$ yr$^{-1}$ even towards highly extincted environments ($\Av\approx50$~mag).

\end{abstract}


\section{Introduction} \label{sec:intro}
In the past decades, technological advances provided the opportunity to study protoplanetary disks, the birth place of planets, in great detail. Radio observations with the Atacama Large sub-Millimeter Array (ALMA) and Near-Infrared (NIR) observations with high-contrast instruments like SPHERE \citep{Beuzit2008} and GPI \citep{Macintosh2008} revealed a plethora of structures in these disks, in the form of rings and gaps, spirals, crescents, and other forms of features \citep[e.g.][]{Andrews2018, Avenhaus2018, Oberg2021}. Even though other explanations are possible, it is believed that at least some of these structures are the result of the gravitational interaction with forming protoplanets (see \citealt{Bae2022_pp7} for a recent review). Consequently, many of these disks have been the target of direct imaging campaigns in the NIR, where we can search for thermal emission from forming planets, and at optical wavelengths, which probe accretion tracers like H$\alpha$. Most of these efforts did not result in protoplanet detections \citep[e.g., ][]{Cugno2019, Zurlo2020, AsensioTorres2021, Huelamo2022, Follette2023, Cugno2023} despite recent detection of a growing number of protoplanets and protoplanet candidates \citep[e.g., PDS70b and c, AB~Aur~b, HD169142~b, AS209~b, MWC758~b and c;][]{Keppler2018, Muller2018, Currie2022, Hammond2023, Bae2022, Reggiani2014, Wagner2023}.

After rings and gaps, spirals are the most common substructure observed in protoplanetary disks \citep{Bae2022_pp7}. Theory predicts two ways to form spirals: 1) if the disk is massive enough they can form via gravitational instability \citep[GI;][]{Rice2003, Lodato2004, Lodato2005, Dong2015-GI}; 2) if a planet is forming in the disk, spirals can form as a result of planet-disk interaction via Lindblad resonances. 
In the second scenario, multiple theoretical works tried to relate spiral properties like their multiplicity, contrast and azimuthal separation to planet characteristics like separation and mass. For example, \cite{Bae2018_mechanism} found that low-mass planets launch a relatively large number of spirals inside their orbits, but if the planet is massive enough ($\Mp\gtrsim 3 M_{\mathrm{th}}$, where $M_\mathrm{th}$ is the thermal mass $M_\mathrm{th}=M_*\times(h/r)^3$ with $h/r$ being the aspect ratio of the disk and $M_*$ being the stellar mass) only two pairs of spirals form as the others merge with the primary and secondary (each pair formed by an inward and an outward spiral). Regarding the spiral contrast, \cite{DongFung2017} conducted 3D hydrodynamical simulations trying to relate the planetary mass responsible for launching the spirals with the measured brightness with respect to the disk, although this method can be fooled by other mechanisms like vortices or shadows and should be used with caution \citep{Bae2018_implications}. Finally, there is a general consensus that more massive planets generate spirals with larger azimuthal separations, until the planet-star mass ratio approaches $\sim0.01$, at which point the primary and secondary spirals move into an $m=2$ symmetry \citep{FungDong2015, Zhu2015, Lee2016, Bae2018_implications}. 

All these works allow to make a prediction of the planet mass perturbing the disk and generating the spirals that can be readily tested with high-contrast imaging methods.
Among the available telescopes, JWST is expected to substantially contribute to the detection and characterization of forming planets thanks to its unique infrared sensitivity and its spatial resolution. Among the several programs targeting Young Stellar Objects (YSOs), the NIRCam Guaranteed Time Observation (GTO) program PID 1179 using NIRCam imaging \citep{Rieke2023} targeted five protoplanetary disks (HL~Tau, MWC758, SAO~206462, PDS70 and TW Hya) with previously observed features and structures hinting at ongoing planet formation. A parallel NIRISS GTO program PID 1242 using NIRISS Aperture Masking Interferometry \citep{Sivaramakrishnan2023} targeted three protoplanetary disks (SAO~206462, PDS70, and HD100546). In this paper, we present the results of the NIRCam observations targeting SAO~206462, while in a companion paper Wagner et al, subm. present the MWC758 data.  
In Sect.~\ref{sec:target}, we provide information on the target and the rationale for inclusion in this program. In Sect.~\ref{sec:observations} we describe the observations, while the data reduction is detailed in Sect.~\ref{sec:data_reduction}. The main results are presented in Sect.~\ref{sec:results} and discussed in Sect.~\ref{sec:discussion}. We draw our conclusions in Sect.~\ref{sec:conclusions}.

\section{SAO~206462}\label{sec:target}
SAO~206462, also known as HD~135344~B, is a F4Ve star \citep{Dunkin1997} located in the Upper Centaurus Lupus star forming region \citep{vanBoekel2005} at a distance of $135.0\pm0.4$~pc \citep{Gaia2022}. It is part of a visual binary system with separation $21^{\prime\prime}$ between the two stars \citep{Mason2001}, and it is surrounded by a transition disk. The most important stellar and disk properties used in this work are summarized in Table~\ref{tab:SAO206462}. 

\begin{table}[t!]
\centering
\caption{Main parameters of SAO~206462 and its disk.}
\def\arraystretch{1.25}
\begin{tabular}{lcc}\hline
Parameter & Value & Reference  \\ \hline
RA (J2000)  & 15 15 48.446 & \cite{Gaia2022}                    \\
DEC (J2000)  & -37 09 16.024  & \cite{Gaia2022}     \\
$\mu_\alpha$ [mas~yr$^{-1}$]  & $-19.21\pm0.03$  & \cite{Gaia2022}  \\
$\mu_\delta$ [mas~yr$^{-1}$]  & $-23.27\pm0.02$  & \cite{Gaia2022}  \\
D [pc]  & $135\pm0.4$  & \cite{Gaia2022}  \\
Spectral Type  & F4Ve & \cite{Dunkin1997} \\
$\Mstar$ [$\Msun$]  & $1.6\pm0.1$ & \cite{Garufi2018}  \\
Age [Myr]  & $11.9^{+3.7}_{-5.8}$ & \cite{Garufi2018}  \\
\multirow{2}{*}{disk $i$ [$^\circ$] $^\mathrm{(a)}$} & \multirow{2}{*}{$13.5\pm2.5$} & \cite{Perez2014} \\
 &  & \cite{vanderMarel2016} \\
disk PA [$^\circ$] $^\mathrm{(b)}$& $62$ & \cite{Perez2014} \\ \hline
\end{tabular}\\
\tablenotetext{\mathrm{a}}{We note that the disk inclination measurements have not yet converged to a definitive value. Here we consider a value of $13.5\pm2.5$ that encompasses multiple values from the literature} \citep[e.g.,][]{Perez2014, vanderMarel2016}. 
\tablenotetext{\mathrm{b}}{No errorbar provided. }
\label{tab:SAO206462}
\end{table}

\subsection{Disk observations}

\cite{Brown2007} studied the SED of the star and inferred the presence of a disk with a dust cavity at $\sim45$~au.
The first resolved detection of the disk surrounding the star was presented in \cite{Grady2009} and was obtained with HST data. HST only detected the outer regions of the disk, without revealing any substructures. Subsequent polarimetric observations with Subaru/HiCIAO \citep{Muto2012}, VLT/NaCo \citep{Garufi2013} and VLT/SPHERE \citep{Stolker2016} revealed a 28~au small dust grain cavity and two spiral arms extending to $0\farcs6$ \citep[$\sim80$~au,][]{Stolker2016}. The spirals show a brighter peak south-west from the star associated with an emission clump \citep{Bae2016}. Furthermore, the scattered light images presented in \cite{Stolker2016} reveal shadow features that the authors associated to a misaligned inner disk. Even though the analysis of \cite{Bohn2022} did not confirm this hypothesis, the authors indicate that the inner disk morphology of SAO~206462 is likely very complex and could not be captured by their model. 

The cavity detected in scattered light was also identified by ALMA continuum observations \citep{Perez2014, Pinilla2015, Francis2020} tracing sub-mm and mm pebbles, but with a larger size of 48~au. The difference in cavity size can be explained by spatial segregation of mm-sized dust grains \citep{Garufi2013}. The gas in the disk has been studied by \cite{vanderMarel2016}, who found that the cavity in $^{13}$CO and C$^{18}$O is smaller in size, $\sim30$~au, and the $^{12}$CO surface density drop at the cavity edge is 4 orders of magnitude. In addition to the inner cavity, ALMA continuum images detected two rings at 52 and 80~au connected by a filament coincident with the Southern spiral and possibly tracing a planetary wake crossing the dust gap \citep{Casassus2021}. 

In recent years, it has been proposed that monitoring of the spiral motion can provide constraints on its origin (GI, planet-disk interaction) and, in case of the latter scenario, of the separation of the perturber \citep[see ][for an example applied to MWC758]{Ren2020}. \cite{Xie2021} studied the motion of the spirals in scattered light over a 1 year period, trying to determine if they are comoving or if they have distinct dynamics. They found that the spiral rotation rate is not consistent with the gravitational instability scenario, as they move at sub-keplerian velocity with respect to keplerian material at their location. Instead, the velocities suggest the spirals move independently from each other ($3\sigma$ confidence) and the eastern one is consistent with being launched by a companion orbiting with a period of $1130\pm780$~yr (corresponding to a semimajor axis of $123^{+63}_{-18}$~au\footnote{To obtain these values we used the azimuthal velocity measured by \cite{Xie2021} and propagated its uncertainty assuming the noise follows a Gaussian distribution}) assuming a circular orbit coplanar with the disk. Notably, this method does not provide constraints on the planet mass. 

Another key result from analyzing the dynamics of SAO~206462 is that the spirals should not have formed due to gravitational instability: \cite{Xie2021} estimated that in order for GI to be responsible for the spirals, the central object mass would need to be $0.1^{+0.08}_{-0.05}~\Msun$, inconsistent with the mass of the star ($1.6\pm0.1$, \citealt{Garufi2018}). Another method to assess the disk potential for GI relies on estimating the total disk mass.
Assuming optically thin emission, the observed millimeter continuum flux from the disk corresponds to a dust mass of 136 $M_\oplus$, or a total-disk-to-star-mass ratio of 2.5\% assuming a gas-to-dust ratio of 100 \citep{dong18spiral}. This is about an order of magnitude lower than needed for GI to produce prominent two-arm spirals \citep[e.g.,][]{Dong2015-GI}.
Thus the disk is unlikely to be GI unstable, unless the millimeter emission is significantly optically thick, and/or the gas-to-dust ratio is substantially above 100.

\begin{table*}[t!]
\centering
\caption{Summary of observations.}
\def\arraystretch{1.25}
\begin{tabular}{lllllllllllll}\hline
Target & Prog. ID & Filter & $\lambda_\mathrm{pivot}$ & W$_\mathrm{eff}$\tablenotemark{a} & Readout & Subarray\tablenotemark{b} & $N_\mathrm{gr}$ & $N_\mathrm{int}$ & $N_\mathrm{dither}$ & $N_\mathrm{roll}$\tablenotemark{c} & $t_\mathrm{tot}$  & FWHM\tablenotemark{d} \\
& & & ($\mu$m)  & ($\mu$m) & & & & & & & (s) & ($^{\prime\prime}$) \\ \hline
SAO~206462 & 1179 & F187N  & 1.874 & 0.024 & RAPID & SUB160 & 10 & 120 & 4 & 2 & 2674 & $0\farcs064$      \\
SAO~206462 & 1179 & F200W  & 1.990 & 0.461 & RAPID & SUB160 & 10 & 120 & 4 & 2 & 2674 & $0\farcs066$      \\
SAO~206462 & 1179 & F405N  & 4.055 & 0.046 & RAPID & SUB160 & 10 & 120 & 4 & 2 & 2674 & $0\farcs136$   \\
SAO~206462 & 1179 & F410M  & 4.092 & 0.436 & RAPID & SUB160 & 10 & 120 & 4 & 2 & 2674 & $0\farcs137$     \\
P330-E    & 1538 & F187N  & 1.874 & 0.024 & RAPID & SUB160 & 7  & 2 & 4 & 1 & 15.6   & $0\farcs064$       \\
P330-E    & 1538 & F200W  & 1.990 & 0.461 & RAPID & SUB160 & 3  & 2 & 4 & 1 & 6.7    & $0\farcs066$      \\
P330-E    & 1538 & F405N  & 4.055 & 0.046 & RAPID & SUB160 & 10 & 2 & 4 & 1 & 22.3   & $0\farcs136$   \\
P330-E    & 1538 & F410M  & 4.092 & 0.436 & RAPID & SUB160 & 3  & 2 & 4 & 1 & 6.7    & $0\farcs137$     \\ \hline
\end{tabular}\\\vspace{0.2cm}
\tablenotetext{a}{Filter bandwidth, defined as the integral of the normalized transmission curve \footnote{\url{https://jwst-docs.stsci.edu/jwst-near-infrared-camera/nircam-instrumentation/nircam-filters}}.}
\tablenotetext{b}{SUB160 has a group time of $t_\mathrm{gr}$ of 0.27864 s. Each integration length was $t_\mathrm{gr} \times N_\mathrm{gr}$.}
\tablenotetext{c}{Number of spacecraft rolls in which the observations is repeated.}
\tablenotetext{d}{Empirical PSF FWHM provided in JWST documentation \footnote{\url{https://jwst-docs.stsci.edu/jwst-near-infrared-camera/nircam-performance/nircam-point-spread-functions}}.}
\label{tab:observations}
\end{table*}

\subsection{Modeling efforts}
\label{sec:HD135_theory}

Hydrodynamical simulations of the SAO~206462 disk allow constraints on the mass and location of the planet generating the detected spirals. Using their scaling relation between star-planet mass ratio $q$ and azimuthal separation of the primary and secondary spirals $\phi_\mathrm{sep}$, \cite{FungDong2015} found the planet mass to be $5.4\pm1.6~\MJ$, where we used the updated stellar mass from Table~\ref{tab:SAO206462}. \cite{Bae2016} performed 2D two-fluids hydrodynamical simulations, confirming that a gas giant planet could be responsible for the observations of both the spiral arms in scattered light and the asymmetric thermal emission observed in ALMA continuum. They inferred the mass of the perturber to be of the order of $10-15~\MJ$ to reproduce the observed features. Finally, \cite{DongFung2017} used the contrast ratio of the spirals to constrain the planet mass, finding that $\Mp\sim5-10~\MJ$ (assuming a planet at $\sim100$~au). In addition, the presence of only two spirals suggest that $\Mp\gtrsim5.1~\MJ$ \citep{Bae2018_mechanism, Bae2018_implications}. All these works agree that the necessary planetary mass to launch both spirals must be at least $\sim5~\MJ$ at $100-120$~au. Assuming the spiral-driving companion is on a circular and coplanar orbit, and the disk has reached a steady state response to the companion (i.e., the companion did not form in the recent past), more massive companions could be located at larger separations from the spirals, because a massive companion opens a bigger gap around its orbit \citep{Dong2016}.

\subsection{Previous high-contrast imaging observations}

SAO~206462 has been observed with the high-contrast imagers GPI, SPHERE and NaCo to search for the thermal emission from planets that are predicted to be responsible for the spirals \citep{Vicente2011, Wahhaj2015, Maire2017, AsensioTorres2021}. So far no gas giant has been detected, and \cite{AsensioTorres2021} reported SPHERE/IRDIS contrast limits ($\Delta K=14$~mag) that exclude planets with $\Mp>4~\MJ$ beyond 100~au when using hot-start models \citep[AMES-DUSTY,][]{Chabrier2000} and $\Mp>11-12~\MJ$ when using warm-start models \citep[BEX-WARM]{Linder2019, Marleau2019} and assuming no extinction affects the planet emission. These were the deepest detection limits available until now.
Another mechanism to search for forming planets is to detect localized H$\alpha$ emission associated with accretion processes onto the planet. SAO~206462 was part of the H$\alpha$ surveys presented in \cite{Zurlo2020}, \cite{Follette2023} and \cite{Cugno2019}. The latter obtained an H$\alpha$ luminosity upper limit of $8.5\times10^{-7}~\Lsun$ with SPHERE/ZIMPOL.

\section{Observations} \label{sec:observations}
\subsection{SAO~206462}
We observed SAO~206462 as part of the NIRCam GTO program (PID 1179, PI: J.\ Leisenring). The program targets five Young Stellar Objects (YSOs) showing strong signposts of ongoing planet formation. The SAO~206462 observations were executed on UT 2023-02-16 in subarray imaging mode (SUB160) using four different filters of the NIRCam instrument \citep{Rieke2023}: F187N, F200W, F405N and F410M (see Table~\ref{tab:observations} for details). The narrow filters are centered on the Pa-$\alpha$ and Br-$\alpha$ emission lines expected to contribute significantly to the protoplanet emitted flux in case it is accreting material, while the medium and wide filters are much broader and focus on detecting thermal continuum emission from a substellar companion. Observations were obtained simultaneously for the F187N/F405N and the F200W/F410M filter combinations, increasing overall observing efficiency. For each filter pair, we observed the target at two roll angles, separated by $\sim10^\circ$, enabling Angular Differential Imaging \citep[ADI;][]{Marois2006}. For each position angle, we placed the star at four different dither positions to mitigate the effect of bad pixels. To minimize wavefront variations, observations with the two rolls were executed one after the other. Indeed a variation of the wavefront could lead to slightly different PSFs, increasing the stellar residuals in the final images and reducing the achieved contrast. No reference star was observed with the intent of performing Reference Differential Imaging (RDI). Table~\ref{tab:observations} reports technical details for the observations. 

\subsection{P330-E}
As described later, the PSF core of the SAO~206462 data was heavily saturated in every filter. Hence, we used the standard G star P330-E to perform photometric calibration of our data. P330-E was observed as part of the NIRCam calibration program (Prog. ID 1538) on 2022-08-29 for all available NIRCam filters and detector combinations with the SUB160 subarray. The star was placed at the 4 corners of the field of view (FoV), but the PSF core necessary for the photometric calibration was always well within the detector. Table~\ref{tab:observations} reports details of these observations as well.

\section{Data reduction} \label{sec:data_reduction}
\subsection{Initial Reduction}

Our data reduction starts with the {\it \_uncal.fits} frames downloaded from the Mikulski Archive for Space Telescope (MAST). Images underwent the standard Stage 1 and Stage 2 of the {\tt jwst} pipeline (version 1.9.4 with crds version 11.16.20) publicly available, which performed all the necessary calibration steps\footnote{\url{https://jwst-pipeline.readthedocs.io/en/latest/}}. Following the directives in \cite{Carter2023}, we switched off dark current correction and we changed the detection threshold of jumps in the ramp fitting to 5. 
After noticing the strong saturation in the central core of the stellar PSFs, we set the parameter {\tt ramp\_fit.suppress\_one\_group} to {\tt False}, in order to minimize the area where no data is available. Despite this intervention, the central $0\farcs2-0\farcs3$ of the images are already saturated within the first read, and therefore those pixels are flagged as NaNs. 

In addition, the {\tt jwst} pipeline flags pixels presenting jumps as NaNs in the final images. 
To remove image NaNs not connected with the PSF core, for each dither position of the two rolls, we substituted those pixels with the median of the PSFs imaged at different locations after being shifted at the same location as the PSF at hand. 
To center the images, we used cross-correlation on a synthetic and perfectly centered PSF generated with {\tt webbpsf} \citep{Perrin2014} using the closest in time Optical Path Difference (OPD) map. As an input for the stellar spectrum, we fit a stellar+blackbody model to the Near- to Mid-Infrared photometries available in VizieR Photometry Viewer, where the blackbody accounts for inner disk excess emission from $\sim3~\mu$m onwards.

\subsection{PSF subtraction}\label{sec:PSF-subtraction}
At this point the data are treated differently when they are used as science and when they are used as reference to remove the stellar PSF from images taken with the other roll angle. Science data are binned every 5 integrations as we found that this does not affect contrast performance, but speeds up every subsequent process. This is not done for references, as more images carry more information and help with the PSF modeling and subtraction. We include a random shift in every reference image, centered around 0 and with standard deviation equal 0.05 pixels. This step is similar to the small grid dither pattern employed during coronagraphic observations\footnote{\url{https://jwst-docs.stsci.edu/jwst-near-infrared-camera/nircam-operations/nircam-dithers-and-mosaics/nircam-subpixel-dithers/nircam-small-grid-dithers}}. It helps minimizing the impact of small pointing error/shifts and the imperfections of the centering algorithm in order to more accurately subtract the central PSF. 
Finally, to deal with the sometimes irregular saturation, we applied a round central mask at the location of the star with radius $r=0\farcs3$ ($0\farcs8$ for F410M). At this point images have sizes of $3\farcs8\times3\farcs8$ and $7\farcs4\times7\farcs4$ at short and long wavelengths respectively.

Then, we applied principal component analysis \citep[PCA;][]{Amara2012, Soummer2012} as performed in PynPoint \citep{Stolker2019} in order to first model the stellar emission in one roll and then subtract it from the other one. Images are then derotated to a common orientation and residuals from both rolls are finally combined. 
We found that at least 7 principal components (PC) need to be removed in order to best reveal the disk structures and residuals from potential planets. Removing larger numbers of PCs does not improve the residuals or the contrast, but, contrary to ground based observations, it does not increase self-subtraction either thanks to the two-rolls observing strategy.

\begin{figure*}[ht!]
\includegraphics[width=1\hsize]{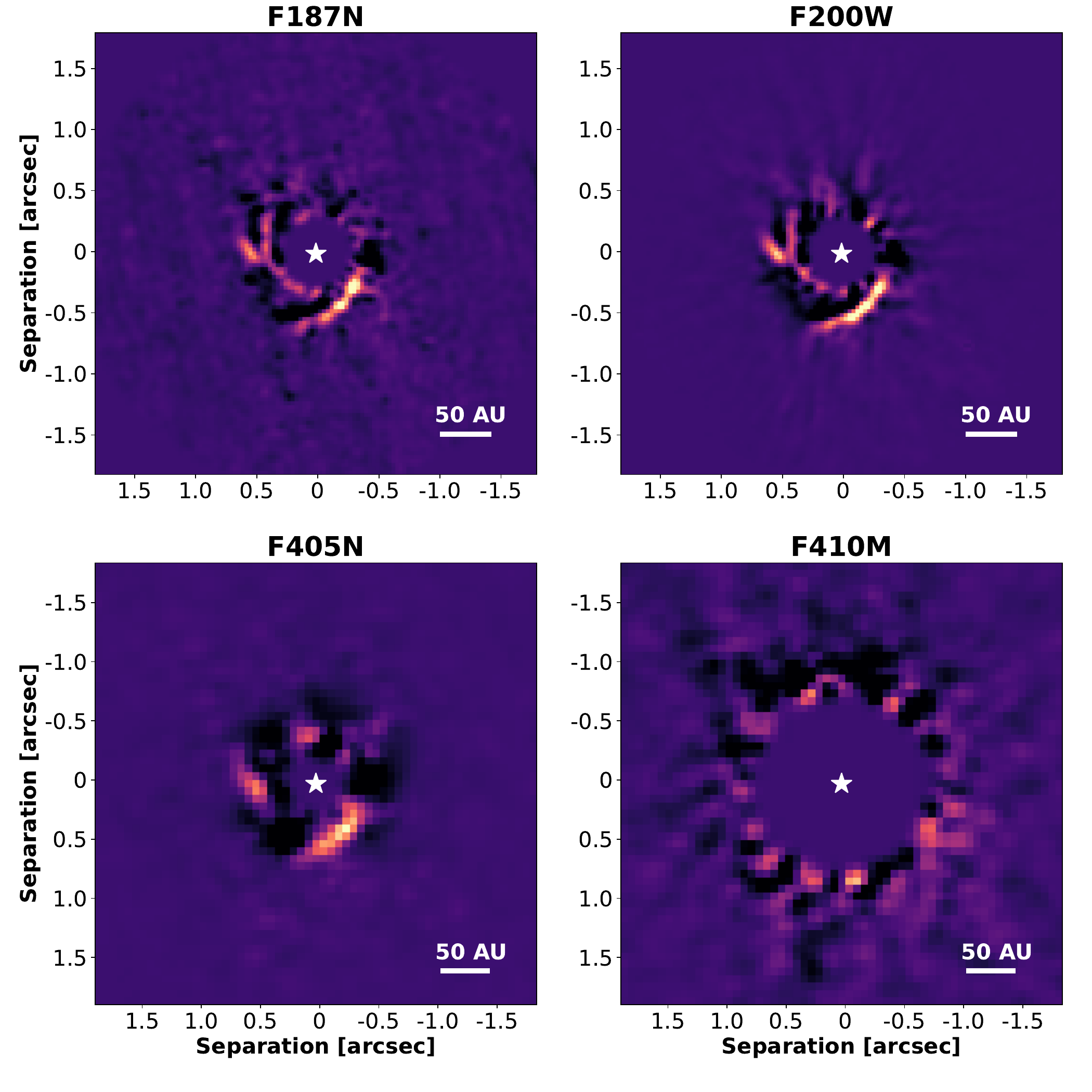}
\caption{SAO~206462 residuals in the four filters observed with our program. The central masks have radii of $0\farcs3$ in F187N, F200W and F405N, and $0\farcs8$ in F410M. In all images North points to the top, East to the left and the field of view is the same. The colorscale has been adapted based on the strength of the disk signal. 
\label{fig:residuals}}
\end{figure*}

\subsection{Photometric PSFs}
The {\it $^*$\_cal.fits} P330-E data were downloaded from the archive and were centered using cross-correlation with a synthetic PSF generated, once again, with {\tt webbpsf}. The images were then median combined, and cropped to $1\farcs0$ in order to focus only on the central region of the PSF. The P330-E spectrum has been taken from Rieke et al., submitted to AJ.

\begin{figure*}[ht!]
\includegraphics[width=1\hsize]{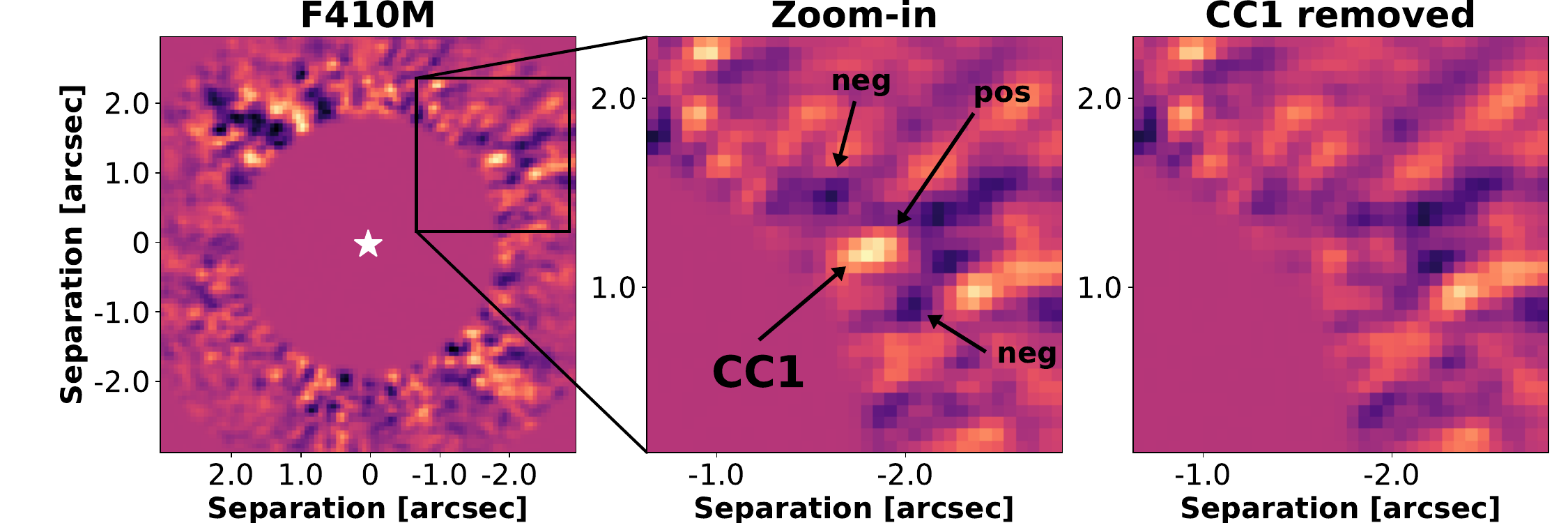}
\caption{F410M residuals highlighting the companion candidate to SAO~206462. {\it Left:} Residuals of the F410M data, similar to Fig.~\ref{fig:residuals}. In this instance we applied a larger central mask to force the PCA to focus on the region of the image around $\sim1\farcs9$. The data reveals a point source in the NW region of the image. North points to the top, East to the left. {\it Center:} Zoomed-in portion of the image, highlighting the candidate. The negative-positive-negative pattern, indicative of a point source with the JWST two-rolls observing strategy is clearly visible. {\it Right:} Same portion of the residuals as in the central panel, but the CC1 pattern was removed injecting one single negative PSF in the data, demonstrating the correlation between the negative and the positive components of the signal. 
\label{fig:candidate}}
\end{figure*}

\section{Results} \label{sec:results}
The PSF-subtracted images for the four filters, shown in Fig.~\ref{fig:residuals}, reveal the circumstellar disk signal in F187N, F200W and F405N, while no disk is detected in F410M. This is due to the strong saturation in the image core that made the inspection of the inner regions not feasible. Due to the two-rolls PSF removal strategy, a considerable fraction of the disk intensity is lost in the PSF subtraction process, and it is currently not possible to extract neither a precise morphological structure, nor fluxes to perform color analysis. This is left for future work once high-contrast at very small separation can be achieved with Reference Differential Imaging via synthetic PSFs \citep[see][]{Greenbaum2023}.

The inspection of the residuals at larger separations reveal a companion candidate in F410M that we will discuss in Sect.~\ref{sec:cc1}. No other point-like signal is detected in the images at any separation, in particular closer to the spirals, and in Sect.~\ref{sec:limits_results} and \ref{sec:accretion} we present the detection limits achieved by our observations.

\subsection{A companion candidate at large separations}\label{sec:cc1}
At large separations the residuals in the F410M images revealed a companion candidate at $\sim2\farcs2$ from the star that we designate CC1. The candidate is best visible when using a larger mask to cover the central region of the images, allowing the principal components to focus on removing the PSF pattern at large separations. In Fig.~\ref{fig:candidate} we report the residuals when using a central mask of $1\farcs9$ arcsec (left panel) and a zoom-in on the area of interest (central panel). Despite several speckles having similar strengths as CC1, the negative-positive-negative pattern expected for a point source in observations obtained with the two-rolls strategy is clearly visible in the residuals, strongly supporting a point source detection.
We additionally inspected the residuals from individual rolls. In both of them there is a positive signal at the location of the source, and a negative feature on one of the two sides (as expected), providing additional evidence that the detection is real.
Using the metric proposed by \cite{Mawet2014}, we estimated the signal-to-noise ratio of CC1 to be 4.4 and its False Positive Fraction to be $1.6\times10^{-5}$, corresponding to $4.1\sigma$ assuming Gaussian noise. This estimate only considers the central positive signal, and being able to consider the correlated symmetric pattern would certainly provide a higher confidence. Note that even though other regions in the image have similar brightness (in particular above the North-East edge of the central mask), they are part of an extended diagonal noise feature resulting from PSF spiders residuals and are not physical sources (see Fig~\ref{fig:diag_masked}, where we masked out the extended feature). In App. \ref{sec:fake_planet} we provide examples of injected sources at the same separation as CC1 and with similar intensity, but at different position angles. In most cases the injected sources indeed look very similar to CC1 in Fig.~\ref{fig:candidate}.

To characterize the companion candidate, we adapted the methods used in \cite{Stolker2020_miracles} to the NIRCam data structure to determine three parameters: separation, position angle and contrast. Briefly, we used an MCMC algorithm to explore the parameter space with 200 walkers performing a chain of 400 steps each. At each step, artificial negative copies of the PSF (obtained from the P330-E data) were inserted in the data at the location of companion candidates. Then, PSF is removed with the same approach described in Sect.~\ref{sec:PSF-subtraction} and the residuals are minimized in an aperture of radius $0\farcs3$ \citep{Wertz2017, Stolker2019, Stolker2020_miracles}. 

To avoid measurement biases and explore potential systematics due to remaining PSF and detector artifacts, we inserted artificial planets at the same separation measured from the MCMC algorithm with the same contrast at 360 different position angles. We then used the same algorithm to retrieve these values and calculate the difference with the input for the 360 experiments. The mean of the differences is added to the MCMC results as a bias offset, while the standard deviation is used to account for speckle noise and systematics in the measurement and is added to the MCMC uncertainties. 

We found that the separation from the central star is $2.225^{+0.071}_{-0.073}$ arcsec and its position angle is $303.5^{+2.0}_{-2.1}$$^\circ$. Considering the disk geometry (Table~\ref{tab:SAO206462}) and assuming the candidate is a real companion coplanar with the circumstellar disk, we find its physical separation to be $300.8^{+9.9}_{-9.5}$~au. 

The measured companion photometry in F410M is $1.7^{+0.6}_{-0.5}~\mu$Jy. Assuming unobscured photospheric emission, we used the BEX evolutionary models \citep{Linder2019} coupled with the HELIOS \citep{Malik2017} atmospheric models assuming solar metallicity, to estimate the mass of CC1. We find $M_\mathrm{CC1}=0.8\pm0.3~\MJ$. This value is consistent with the non-detection in F200W (see below). We note that inserting one negative point source in the data at the CC1 location removes both the negative wings and the central positive signal (see right panel of Fig.~\ref{fig:candidate}), confirming once again the correlation between the elements of the signal pattern and excluding independent speckles residuals could generate the CC1 signal.

The point source is not detected in the F405N filter, as the LW narrowband data are less sensitive in that region than the F410M data (see Sect.~\ref{sec:limits_results}). Furthermore, due to the SW data's smaller pixel scale, the location of CC1 falls outside of the FoV for some of the dither positions. Hence, we used only two dither positions for each roll where the star is located such that the location of CC1 is included in the frames. CC1 is not detected in the final residuals, and using the approach described in the next Section we estimated a flux upper limit of $0.8~\mu$Jy in the F200W filter at the separation of CC1. This implies that CC1 is a very red object. 

To estimate the probability of CC1 being a background object within the Milky Way galaxy, we used the Trilegal galactic model \citep{Girardi2012}, which provides, given the coordinates, a model for the stellar population observable in an area of 1~deg$^2$. To each element of the catalogue consistent with the F200W limits, we assign a value corresponding to the probability function of the CC1 F410M photometry. We then sum up these values and normalized to the considered FoV (we used 55~arcsec$^2$ for the NIRCam data, equivalent to the FoV of the F410M data). 
This translates in a probability of 0.008 that CC1 is a background galactic object. 

Additionally, we used the mock extragalactic catalog from \cite{Williams2018} to estimate the probability of CC1 being a background extragalactic object. The catalog includes two populations of galaxies (quiescent and star-forming) and both need to be considered. Given the position of SAO~206462 in the Milky Way (latitude 332.392$^\circ$; longitude +17.325$^\circ$), galactic extinction ($\Av=0.15$~mag\footnote{Value taken from \url{https://irsa.ipac.caltech.edu/cgi-bin/bgTools/nph-bgExec}}) towards a background object can be considered negligible, especially in the infrared. We found that for the star-forming and quiescent galaxy populations CC1 has a probability of 0.354 and 0.001 of being a background contaminant respectively. 

\begin{figure}[b!]
\includegraphics[width=\hsize]{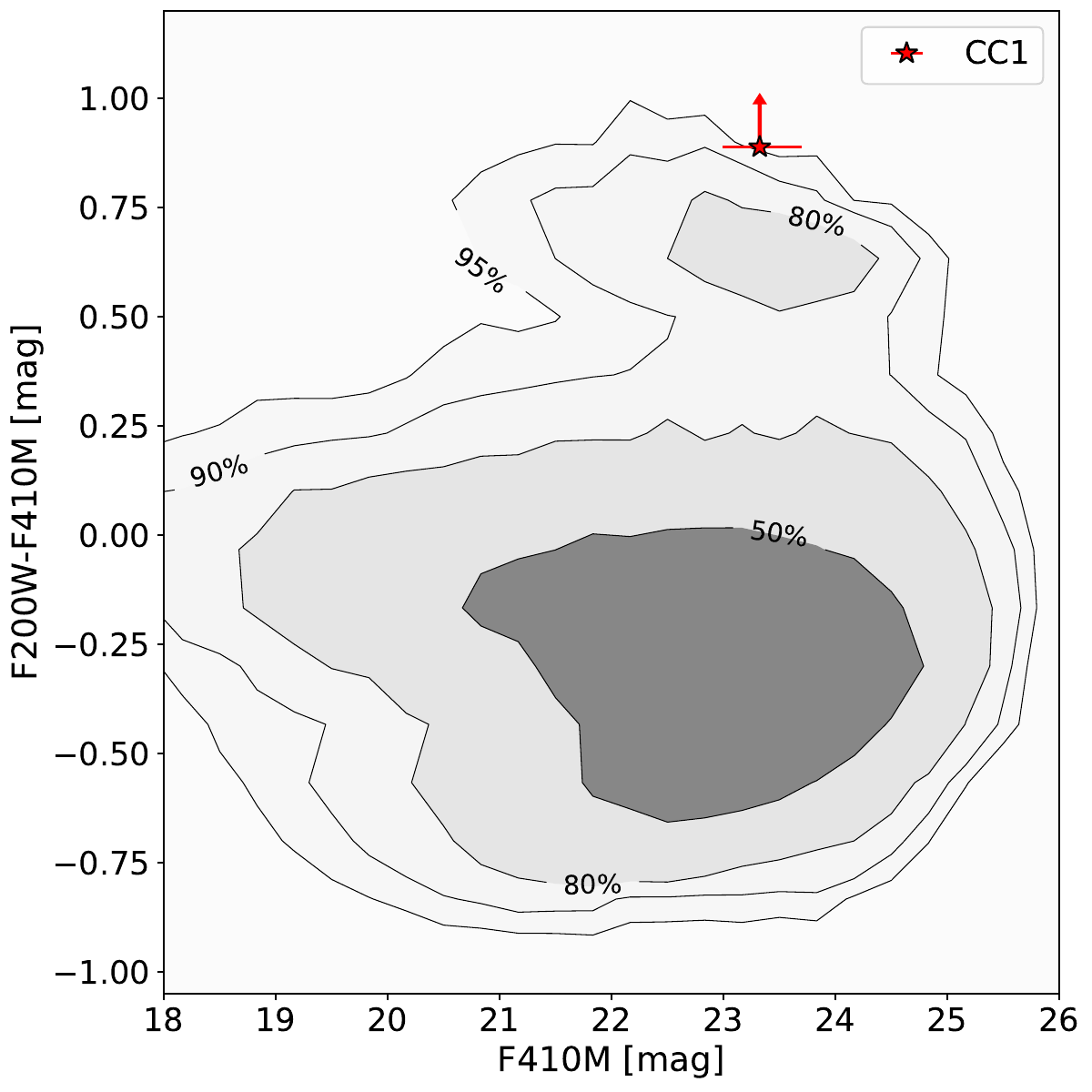}
\caption{ Color-magnitude diagram of CC1 and the galaxy population from the JAGUAR mock catalog \citep{Williams2018}. Black lines represent the contours including 50\%, 80\%, 90\% and 95\% of the galaxies in the catalog. CC1 is located at the edge of the galaxy population due to its very red colors. 
\label{fig:cmd}}
\end{figure}

The location of CC1 in a color-magnitude diagram (CMD) is shown in Fig.~\ref{fig:cmd}. In grey, we show the population of galaxies with the contours indicating the areas including 50\%, 80\%, 90\% and 95\% of the elements. The very red colors of CC1 are inconsistent with the vast majority of the extragalactic objects. 

Combining results from the galactic and extragalactic models, there is a probability of $\sim36.3\%$ that CC1 is a background object based on the NIRCam observations. This value should be treated as an upper limit, as many of the background contaminant galaxies from the mock catalog are spatially resolved, while CC1 is unresolved. To include the requirement that the contaminant should be unresolved, we can take the effective radius $r_e$ of each source and convolve it with the FWHM of the F410M PSF (see Table~\ref{tab:observations}. In practical terms, we verify that the term $\sqrt{r_e^2 + \mathrm{FWHM}_\mathrm{F410M}^2}$ is smaller than a threshold that we set to $1.5 \mathrm{FWHM}_\mathrm{F410M}$. This reduces the probability of the contaminant scenario to $\sim4.2\%$ (galactic and extragalactic). Follow-up observations in future cycles will help determining the true nature of CC1.

\subsection{Sensitivity limits}\label{sec:limits_results}
To estimate the contrast limits we used {\tt applefy} \citep{Bonse2023}, a python package that estimates contrast curves by injecting artificial sources in the data in steps of 1 FWHM (see Table~\ref{tab:observations}), where at each separation we inspected 6 different azimuthal positions. We set the detection threshold to a False Positive Fraction (FPF) of $2.87\times10^{-7}$, corresponding to $5\sigma$ in the case of Gaussian noise. The noise was estimated as individual pixels separated by 1 FWHM \citep{Bonse2023}. To take into account the speckles field in the residuals, at each separation the noise configuration was measured 360 different times, each with slightly different angular orientation, using the obtained median value for the contrast and the standard deviation as an uncertainty on the contrast measurement. 

Due to the central saturation in our images, we used PSFs from P330-E (see Sect.~\ref{sec:observations}) to inject fake companions. This choice prevents us from obtaining contrast estimates compared to the central star, but guarantees an accurate photometric calibration of our limits. Indeed, the contrast estimated with respect to the P330-E PSF can be directly combined with the flux of the standard star (20.17, 21.23, 5.81 and 5.84 mJy for F187N, F200W, F405N and F410M, respectively) in order to have a calibrated flux limit.

\begin{figure}[ht!]
\includegraphics[width=\hsize]{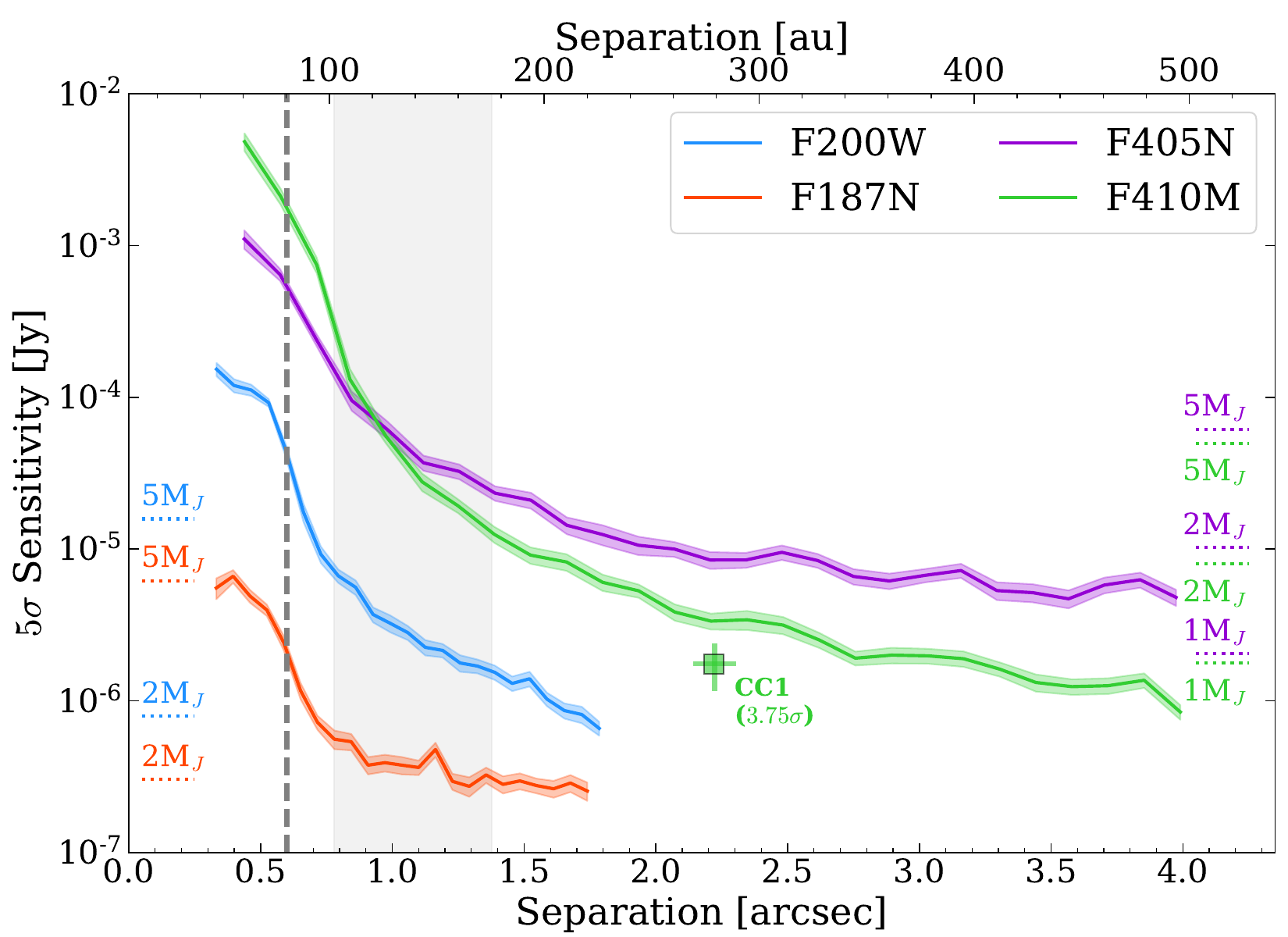}
\caption{NIRCam 5$\sigma$ flux sensitivity limits for SAO~206462 as a function of separation. The dashed vertical line represents the size of the scattered light disk observed in our data, while the gray shaded area highlights the area around $120\pm30$~au where \cite{Xie2021} predicted a planet could be located. The green square represents the measured flux for CC1 in the F410M filter detected with a confidence of $\sim3.8\sigma$. On the left (right) side of the plot, we marked the flux expected in the SW (LW) filters for a 1,2 and 5~$\MJ$\ object at $\sim12$ Myr according to the BEX evolutionary model. 
\label{fig:limits}}
\end{figure}

Following this procedure we estimated the 5$\sigma$ contrast limits for our SAO~206462 NIRCam datasets compared to P330-E. From there, we used the calibrated fluxes to convert the contrast measurements into flux limits, which are reported in Fig.~\ref{fig:limits} and Table~\ref{tab:sensitivity}. The gray dashed vertical line represents the size of the spirals detected in scattered light (consistent with \citealt{Stolker2016}), and therefore marks the region where the limits can not be trusted. Indeed, at shorter separations the physical signal from the disk and the strong self-subtraction prevent the noise from being independent and identically distributed, and the statistics should be treated with caution \citep{Cugno2023}. 
For the SW filters, the narrow filter reaches a lower sensitivity than the broad filter. This is likely because only the inner $0\farcs8$ of the F187N data are contrast dominated, while at larger separation the data are background limited. Conversely, F200W images are contrast limited at every separation, and the sensitivity limit of the filter is expected to be almost two orders of magnitude lower than what we achieve at large separations. In the LW channel, F405N is more sensitive $<0\farcs9$, while at larger separation F410M could detect fainter sources. Indeed, at short separations the effect of saturation and its consequences like the brighter-fatter effect and charge migration play an important role in limiting our sensitivity in F410M. Both LW filters are far away from being background limited, even at large separations. 

\begin{figure*}[ht!]
\centering
\includegraphics[width=0.49\hsize]{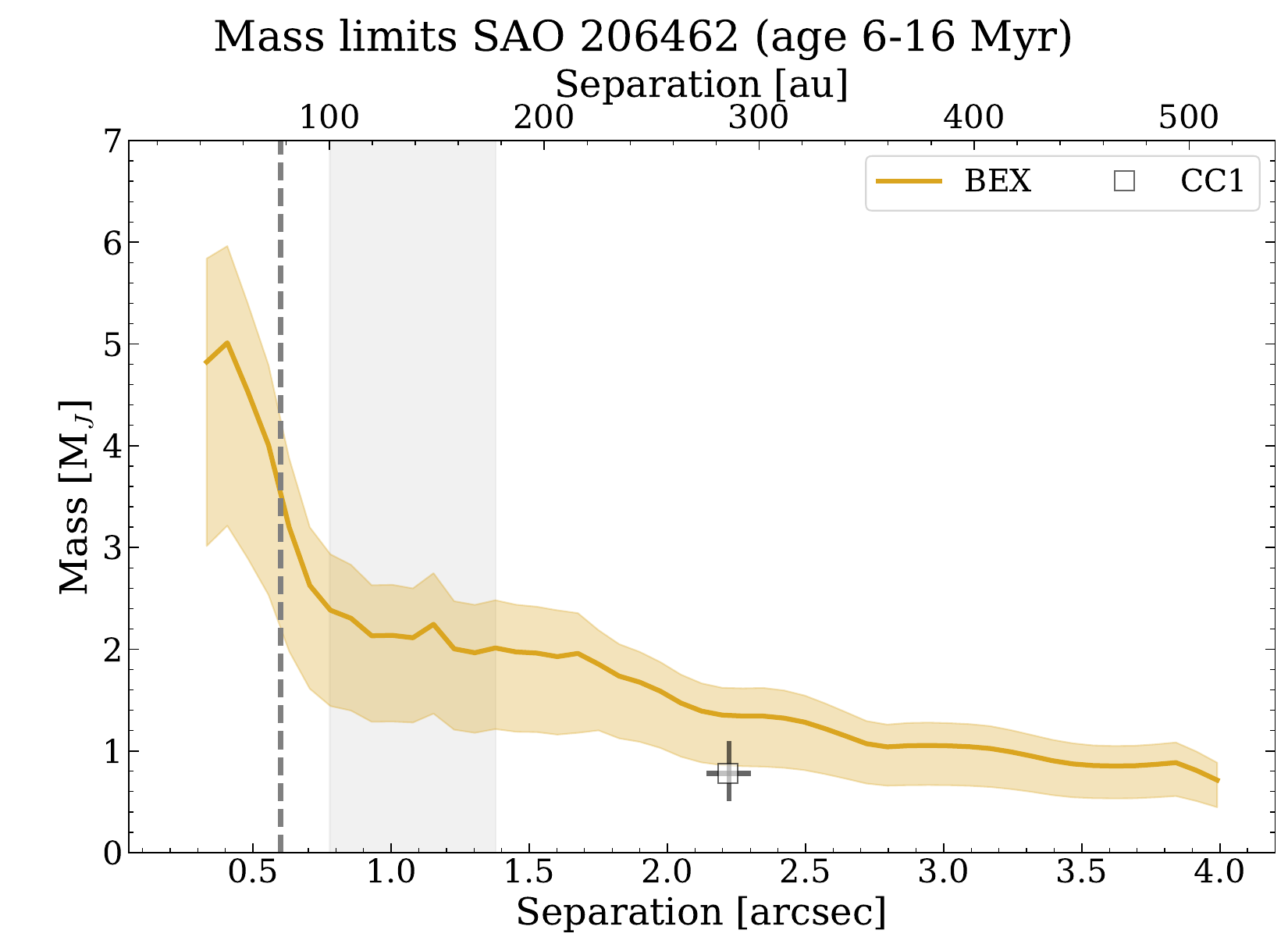}
\includegraphics[width=0.49\hsize]{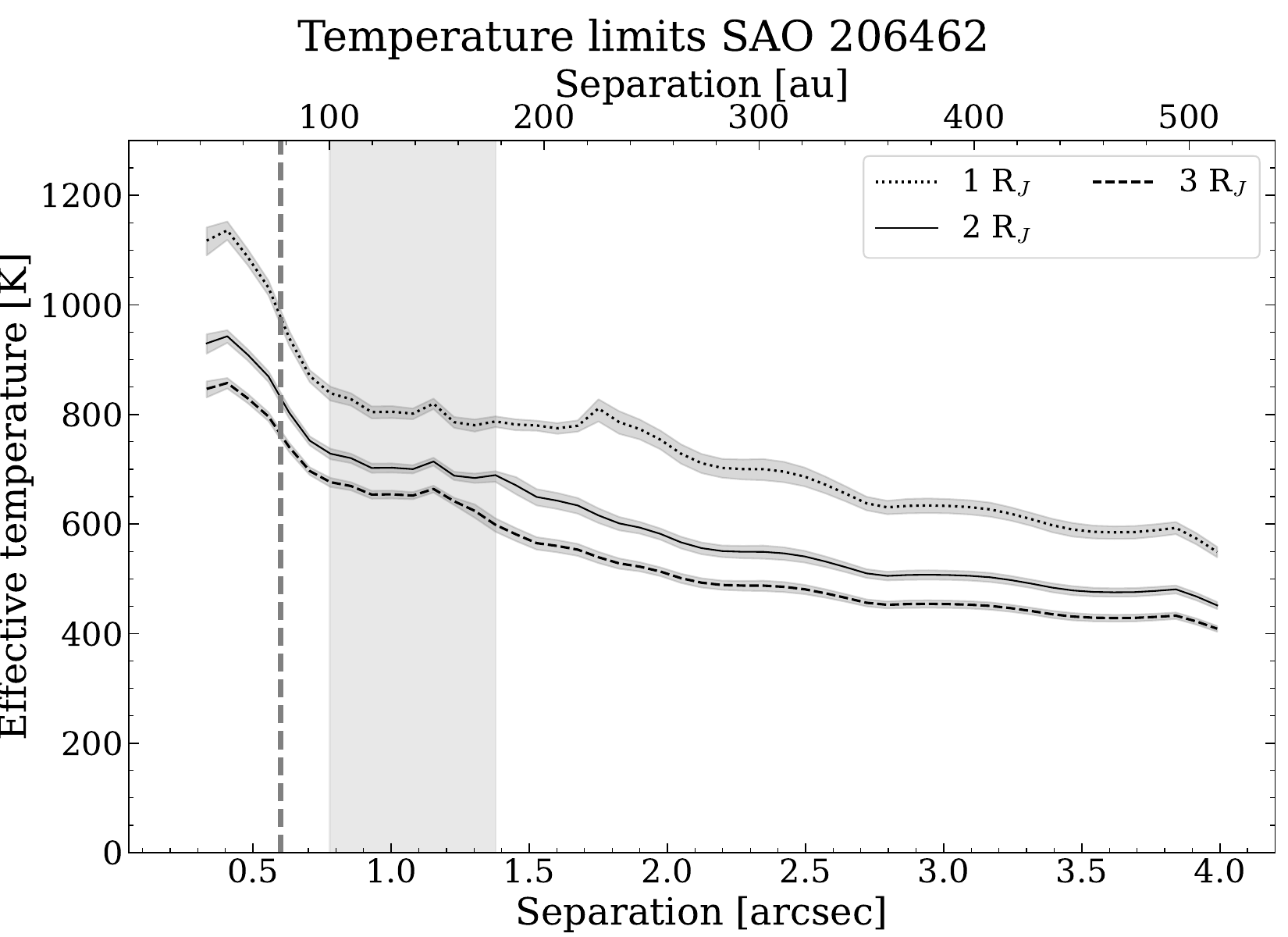}
\caption{Mass and temperature limits of SAO~206462 as a function of separation obtained from the 5$\sigma$ flux limits. The dashed vertical line represents the size of spirals in scattered light and the shaded area highlights the predicted planet location from \cite{Xie2021}. {\it Left:} Mass upper limit when using the BEX cooling curve for the age of the system. The uncertainties, represented by the shaded area, include the age uncertainties of the system ($6-16$ Myr). The estimated mass of CC1 is showed as a black square. At $\sim120$~au, the NIRCam data exclude planets with masses $\gtrsim 2.2~\MJ$. {\it Right:} Effective temperature 5$\sigma$ upper limits when assuming protoplanets emit as blackbodies with radii 1, 2, and 3 $\RJ$. 
\label{fig:mass_teff_limits}}
\end{figure*}

We used the same evolutionary models from Sect.~\ref{sec:cc1} to interpret the sensitivity limits. At each separation, we converted fluxes into planet masses for the four filters and then considered only the most constraining mass limit achieved. The final mass limits can be found on the left panel of Fig.~\ref{fig:mass_teff_limits} and Table~\ref{tab:sensitivity}. At 120~au, the NIRCam data exclude planets more massive than $2.2^{+0.5}_{-0.9}~\MJ$ for a system age of $11.9^{+3.7}_{-5.8}$~Myr, while at larger separations ($\gtrsim2\farcs0$) we would have detected objects with $M>1.5^{+0.3}_{-0.5}~\MJ$, assuming negligible extinction. These mass detection limits are much deeper than those previously available through VLT/SPHERE \citep{AsensioTorres2021}.

In addition to the model-based mass estimates, we want to pursue a more empirical approach. Following \cite{Cugno2023}, we assumed that forming planets emit like blackbodies. This assumption is consistent with observations of PDS70~b and c \citep{Wang2020, Stolker2020, Cugno2021} where no significant molecular features have been detected so far\footnote{We note that the VLTI/GRAVITY spectrum presented in \cite{Wang2021} shows a small dip in the $K$-band spectrum potentially connected to a water feature, but a confirmation is still lacking.}. We fixed the planet radii to 1, 2 and 3 $\RJ$ representing a range of radii consistent with planet formation and evolutionary models \citep[e.g.,][]{Allard2001, Baraffe2003, GinzburgChiang2019} or observations \citep{Stolker2020}. We then explore $\Teff$ to find the maximum value that produces a blackbody with emission within the NIRCam filters below our limits. 
The $\Teff$ limits as a function of separation are showed in the right panel of Fig.~\ref{fig:mass_teff_limits} and are reported in Table~\ref{tab:sensitivity}. At the predicted separation of the planet \citep{Xie2021} our data are sensitive to objects with $\Teff\gtrsim650-830$~K (depending on $\Rp$), while beyond $2\farcs0$ we hit the $500-750$~K limit.

\subsection{Accretion processes}
\label{sec:accretion}
The two narrow filters used in this work allow to investigate the presence of ongoing planetary accretion processes, as they are centered on two hydrogen recombination lines (Pa-$\alpha$ and Br-$\alpha$). In this section, we will consider the limits estimated at 120~au (see Fig.~\ref{fig:limits}), which provide line luminosity upper limits of $2.1\times10^{-10}~\Lsun$ at Pa-$\alpha$ ($\lambda=1.87~\mu$m) and $8.8\times10^{-9}~\Lsun$ at Br-$\alpha$ ($\lambda=4.05~\mu$m). We assumed three mass values of 1, 5 and $10~\MJ$, where $5~\MJ$ is the predicted mass of the perturber generating both spirals (see Sect.~\ref{sec:HD135_theory}), $10~\MJ$ represents the upper limit of the possible mass range obtained by hydrodynamical simulations and $1~\MJ$ represents a low mass scenario. For each mass, we estimate the planet radius using the BEX cooling curves. 

Using the accretion shock models from \cite{Aoyama2018, Aoyama2020}, \cite{Aoyama2021} determined the  relation between $\Lline$ and $\Lacc$ for forming protoplanets, which can be used to determine $\Lacc$ and subsequently $\Macc$ via 
\begin{equation}
    \Macc=\frac{\Lacc~\Rp}{G~\Mp}
\end{equation}
where $G$ is the gravitational constant. 

Figure~\ref{fig:macc} reports the mass accretion rate upper limit of the potential planet as a function of the extinction affecting the line emission in case of accretion processes. Following theoretical work on embedded protoplanets, we consider extinction values up to $\Av=150$~mag \citep{Sanchis2020, Szulagyi2019}. The orange and the violet lines report mass accretion rate upper limits based on the new NIRCam data on Pa-$\alpha$ and Br-$\alpha$ respectively presented in this work. In addition, we plot the H$\alpha$ flux upper limit obtained by \cite{Cugno2019} with SPHERE/ZIMPOL ($\LHa= 8.5\times10^{-7}~\Lsun$ at separations $\gtrsim0\farcs25$).

The new JWST limits are much more constraining than the previous ground -based optical observations for every value of extinction. 
For $\Av=0$~mag, the difference between the existing H$\alpha$ limits and the new Pa-$\alpha$ limits in terms of mass accretion rate is $\sim2$ orders of magnitude. The difference increases with increasing extinctions, mainly because of the strong effect of extinction at optical wavelengths. At $\Av\approx40$~mag, JWST is as sensitive at Pa-$\alpha$ as the unobscured ($\Av=0$~mag) H$\alpha$ SPHERE data. Because the effect of extinction continues to weaken with increasing wavelengths, at $\Av\approx60$~mag the strongest constraints are provided by Br-$\alpha$ data, as Pa-$\alpha$ becomes too obscured. Figure~\ref{fig:macc} highlights which filter is more sensitive depending on the obscuration level, and demonstrates the potential of JWST in identifying accreting embedded planets.

\section{Discussion}\label{sec:discussion}

\subsection{What is launching the observed spirals?}

At 120~au, where the spiral dynamics suggest a planet is located \citep{Xie2021}, the NIRCam data are able to exclude objects with fluxes of $4.8\pm0.5$ and $103.7\pm16.6~\mu$Jy in the F200W and F410M filters, respectively (see Fig.~\ref{fig:limits}). When translated into BEX masses, these values exclude planets with $\Mp\gtrsim2.2~\MJ$, at odds with theoretical predictions based on the spirals morphology and brightness, which suggested $\Mp\gtrsim5~\MJ$. When translated into effective temperatures, these flux limits exclude planets hotter than $817\pm11$, $712\pm9$ and $662\pm8$ K assuming a radius of 1, 2 and 3 $\RJ$. Hence, both our $\Mp$ and $\Teff$ limits are not consistent with unobscured hot-start models. For example, AMES-DUSTY and BT-Settl evolutionary models \citep{Chabrier2000, Allard2001, Allard2003} predict that a $5-10~\MJ$ planet at the age of SAO~206462 would have an effective temperature of $1200-1700$~K with a radius of $\sim1.5~\RJ$. Such a companion would be clearly visible in multiple JWST/NIRCam filters. These results point towards two possible scenarios similar to what suggested in \cite{Wagner2023}: either the planet forms cold \citep[e.g.,][]{Marley2007, Spiegel2012}, or it is highly extincted. 

A much cooler evolutionary track would explain many non-detections of predicted planets embedded in disks in the thermal continuum \citep{AsensioTorres2021, Cugno2023} and the NIRCam data provide tight constraints on the possible temperature of the planet around SAO~206462. Cold-start models predict smaller planet radii compared to hot-start models ($\Rp\sim1~\RJ$), and our constraints suggest the planet $\Teff$ should be below 800~K. This is consistent, for example, with the cold-start models presented in \cite{Marley2007} and \cite{Spiegel2012}. We note however, that this scenario is at odds with preliminary results on dynamical masses of giant planets, where it appears that giant planets form warm to hot \citep[e.g.,][]{Brandt2021, Brandt2021_HR8799, Franson2023}. However, two factors must be considered: (i) most planets with dynamical mass measurements are much closer to their star than the planet launching the spirals in SAO~206462, possibly indicating they underwent a different formation pathway or were influenced by different environments at birth; and (ii) this might be an observational bias, as only hot-start planets might be bright enough to be detected by currently available instrumentation. 

In the other scenario, the planet is deeply embedded in the circumplanetary disk/envelope, and the high extinction is hindering its detection at NIR wavelengths. This possibility is supported by observations of the PDS70 system \citep{Isella2019, Benisty2021, Cugno2021} and was already explored by \cite{Wagner2023} for MWC758~c. We estimated that a visual extinction of $\Av\gtrsim20$~mag would be necessary to make sure that a $5~\MJ$ perturber following the BEX models would not appear in any of our NIRCam filters. If obscuration from the circumplanetary material is the main cause for the poor detection rate of protoplanets, longer wavelength observations might be able to trace colder temperatures and reveal the light reprocessed by the planet surroundings.

\subsection{Could CC1 be responsible for the spirals?}\label{sec:discussion_cc1}
When considering the spiral rotation rate from \cite{Xie2021}, CC1's semi-major axis is located at $2.25\sigma$ from the dynamical prediction for the perturber in the eastern spiral. The rotation velocity from \cite{Xie2021} is based on a 1 year baseline, and it is likely that a longer period between observations will provide a tighter and more precise constraint on the separation of the perturber. 
As described in Sect.~\ref{sec:cc1}, assuming its emission is due to photospheric emission from a gravitationally bound object, we estimated the mass of CC1 to be $0.8\pm0.3~\MJ$. It is unlikely that such a small mass at such large separation is responsible for launching both spirals, but if two perturbers are launching the spirals independently, the mass requirement is relaxed and it is possible that a $\sim \MJ$ planet drives one of them.

There are a few other possibilities that could connect the distant CC1 and the two prominent spirals: (i) an eccentric orbit, (ii) a very large mass for CC1 and (iii) a flyby encounter. These scenarios are further discussed in the next paragraphs.

If CC1 is on a highly eccentric orbit ($e\gtrsim0.6$ assuming CC1 is currently at its apocenter and its pericenter is about the disk size), it can be capable of  launching spirals if it is massive enough. \cite{Zhu2022} found that if the companion is on an eccentric orbit, a higher mass is required to form the same spiral pattern. This would imply that the mass of CC1 should be $\gg5~\MJ$. Again, the requirement would be for the object to either start very cold or be highly obscured (or both). We note that
an eccentric orbit could also explain the different spiral velocities measured by \cite{Xie2021}.

If CC1 is highly obscured, it is possible that its mass is much larger than what we found in Sect.~\ref{sec:cc1}. If that is the case we are looking at a system similar to HD100453 \citep{Collins2009, Benisty2017, Wagner2018}, in which a massive (stellar) companion is driving a pair of near-symmetric spirals, which has been shown in simulations as well \citep{Dong2016, Dong2016hd100453}. 
In addition, such a massive companion is expected to open a wide gap \citep{Artymowicz1994},
which can be tested in gas observations.
We note that given the very deep NIRCam measurements at F410M, it is unlikely that CC1 could be as massive as HD100453~B \citep[$0.2\pm0.04~\Msun$,][]{Collins2009}. Indeed a highly embedded stellar companion with properties similar to a Class 0 object would require a dusty envelope that most likely would have been detected in the ALMA data. Hence, the massive companion scenario seems rather unlikely. 

CC1 could be an older (and hence more massive) flyby object able to produce the spiral structure \citep[e.g.,][]{Cuello2019}. \cite{Shuai2022} excluded that a flyby object detected by {\it Gaia} could have induced the spirals, but CC1 does not appear in the {\it Gaia} catalog. Assuming an age of 1 Gyr, we estimate a mass of $30-50~\MJ$ for an object consistent with the F410M measurement. \cite{Smallwood2023} tested perturber masses in the $10~\MJ - 1~\Msun$ range with 3D hydrodynamical simulations, showing that small mass objects produce faint spirals, while more massive bodies are necessary to see prominent features. In this scenario, it is not clear what mass would be required to reproduce the bright spirals of SAO~206462 due to insufficient information on its orbit, and determining this value is beyond the scope of this paper. 

Finally, CC1 could be a faint companion to SAO~206462, but not being the object responsible for lauching the disk spirals.  Such an object is still below our sensitivity and deeper observations at long wavelength are necessary to directly detect it.

\subsection{Constraints on the mass accretion rate}

\begin{figure}[ht!]
\includegraphics[width=1\hsize]{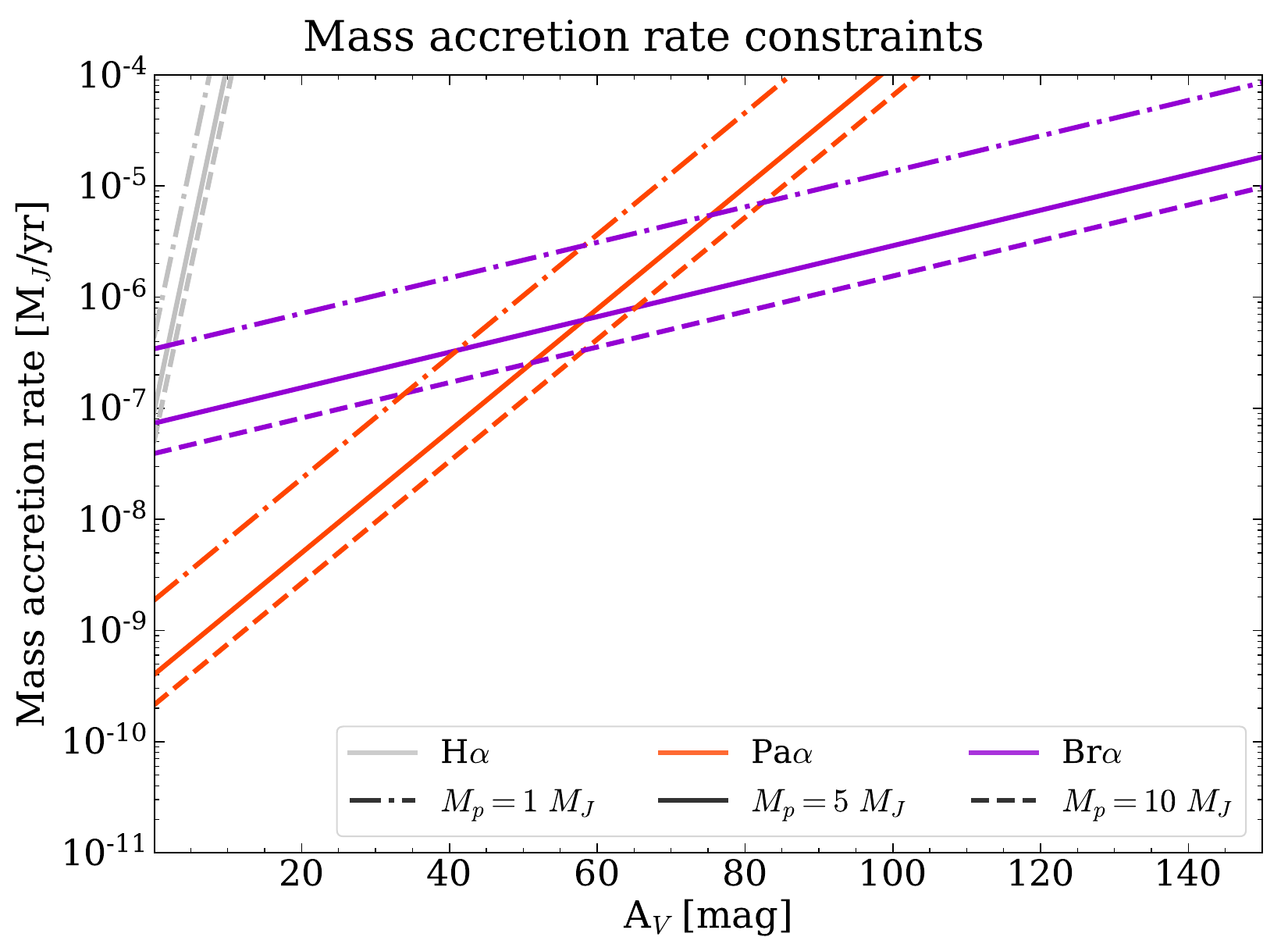}
\caption{Mass accretion rate upper limits as a function of the extinction affecting the lines for planets with masses 1, 5 and 10 $\MJ$. Different colors represent different emission lines and different line styles indicate different planetary masses.
\label{fig:macc}}
\end{figure}

The very tight limits indicate that even for very extincted objects ($\Av=100$~mag), if accretion onto the planet is ongoing at a moderate rate ($10^{-6}~\MJ$~yr$^{-1}$) we should have detected emission in the line filters. For less extincted scenarios ($\Av=40-60$~mag) we exclude even low mass accretion rates of the order $10^{-8}-10^{-7}~\MJ$~yr$^{-1}$ (see Fig.~\ref{fig:macc}). These are the most stringent constraints of ongoing accretion processes in any protoplanetary disk, and we highlight that even the extreme AO systems at H$\alpha$ are not able to achieve such sensitivity \citep{Cugno2023_AS209}. 

The lack of detection of accretion processes despite the extremely deep limits is hard to reconcile with the scenario of a planet strongly interacting with and likely accreting disk material. However, recent radiation hydrodynamics simulations by  \cite{Marleau2023} showed that only a fraction of about 1\% of the accreting material falls onto the planet fast enough to produce a shock and emit hydrogen lines. Alternatively, due to the thick envelope, the incoming material does not shock at the planet surface but at higher elevations, where it does not have enough energy to excite hydrogen atoms and thus produce emission lines. The low emission efficiency and the higher elevation of the shock surface could explain the lack of accretion line detections in the very deep data presented in this work. Further work is warranted both on the theoretical and observational side in order to unveil how young gas giants grow their mass.

\section{Conclusions}\label{sec:conclusions}
In this manuscript, we presented JWST/NIRCam data of SAO~206462, a young star surrounded by a disk with two prominent spirals detected in the NIR. Our data suffered from strong saturation and consequent detector effects, and future observations targeting bright protoplanetary disks should deploy NIRCam coronagraphs to mitigate these issues \citep[][]{Kammerer2022, Girard2022, Carter2023}. Nonetheless, the unparalleled sensitivity of JWST in the $2-4~\mu$m range allowed us to provide tight constraints on the properties of the spiral-launching planet. Our main findings are: 
\begin{itemize}
    \item We detected a companion candidate, CC1 at $2\farcs2$ from the star in the F410M filter. Assuming it is bound to SAO~206462 and not extincted, it has a mass of $0.8\pm0.3~\MJ$. Its semi-major axis is at only $2.25\sigma$ from the predicted separation of the companion driving the eastern spiral \citep{Xie2021}. However, a background contaminant can not be excluded at this point and follow-up observations are necessary to verify common proper motion with SAO~206462.
    \item We did not detect other companions, especially at the separation predicted by previous hydrodynamical simulations. For the BEX models, our limits exclude a planet more massive than $2.2^{+0.5}_{-0.9}~\MJ$ could be located at $\sim120$~au. In terms of temperatures, we exclude objects warmer than $660-810$~K ($1-3~\RJ$). These results suggest the forming planet is either following a cold-start cooling track or its emission is strongly extincted from the surrounding material. 
    \item The accretion line limits at Pa-$\alpha$ and Br-$\alpha$ set tighter constraints on ongoing accretion processes than existing ground-based H$\alpha$ imaging data for every extinction scenario. For high extinction ($\Av>60$~mag), Br-$\alpha$ provides stronger constraints ($\dot{M}\sim10^{-7}-10^{-5} \MJ$ yr$^{-1}$), while for lower extinction the background limited Pa-$\alpha$ data are more sensitive ($\dot{M}\sim10^{-9}-10^{-7} \MJ$ yr$^{-1}$). 
\end{itemize}

Together with the MWC758 data presented in Wagner et al., subm., this paper provides the first JWST/NIRCam high-contrast search for young forming planets embedded in a protoplanetary disk, demonstrating the great capabilities of the new observatory in detecting these objects in the outer regions of circumstellar disks, where data are mostly limited by the background.

\begin{acknowledgments}
We would like to thank Kevin Hainline and Christina Williams for helping in using the JAGUAR models. Furthermore, we would like to thank Nienke van der Marel for verifying the position of CC1 in the ALMA continuum images. We would also like to thank the anonymous referee, whose careful and constructive comments improved the quality of this manuscript.
The authors are grateful for support from NASA through the JWST NIRCam project though contract number NAS5-02105 (M. Rieke, University of Arizona, PI). 
GC thanks the Swiss National Science Foundation for financial support under grant number P500PT\_206785.
D.J.\ is supported by NRC Canada and by an NSERC Discovery Grant.
K.W. acknowledges support from NASA through the NASA Hubble Fellowship grant HST-HF2-51472.001-A awarded by the Space Telescope Science Institute, which is operated by the Association of Universities for Research in Astronomy, Incorporated, under NASA contract NAS5-26555. Some of the data presented in this paper were obtained from the Mikulski Archive for Space Telescopes (MAST) at the Space Telescope Science Institute. The specific observations analyzed can be accessed via \dataset[DOI: 10.17909/65md-vq43]{https://doi.org/10.17909/65md-vq43}. 

\end{acknowledgments}

%

\vspace{5mm}
\facilities{JWST(NIRCam)}


\software{{\tt jwst} version 1.9.4 \citep{jwst_pip}, {\tt PynPoint} \citep{Stolker2019}
          }



\appendix

\section{Masking the diagonal noise residuals}
\label{sec:mask_diag}

In this section we present the same residuals shown in Fig.~\ref{fig:candidate}, where the extended noise feature stemming from the PSF removal is masked. CC1 clearly stands out as the strongest signal in the image. 

\begin{figure}[t!]
\centering
\includegraphics[width=0.5\hsize]{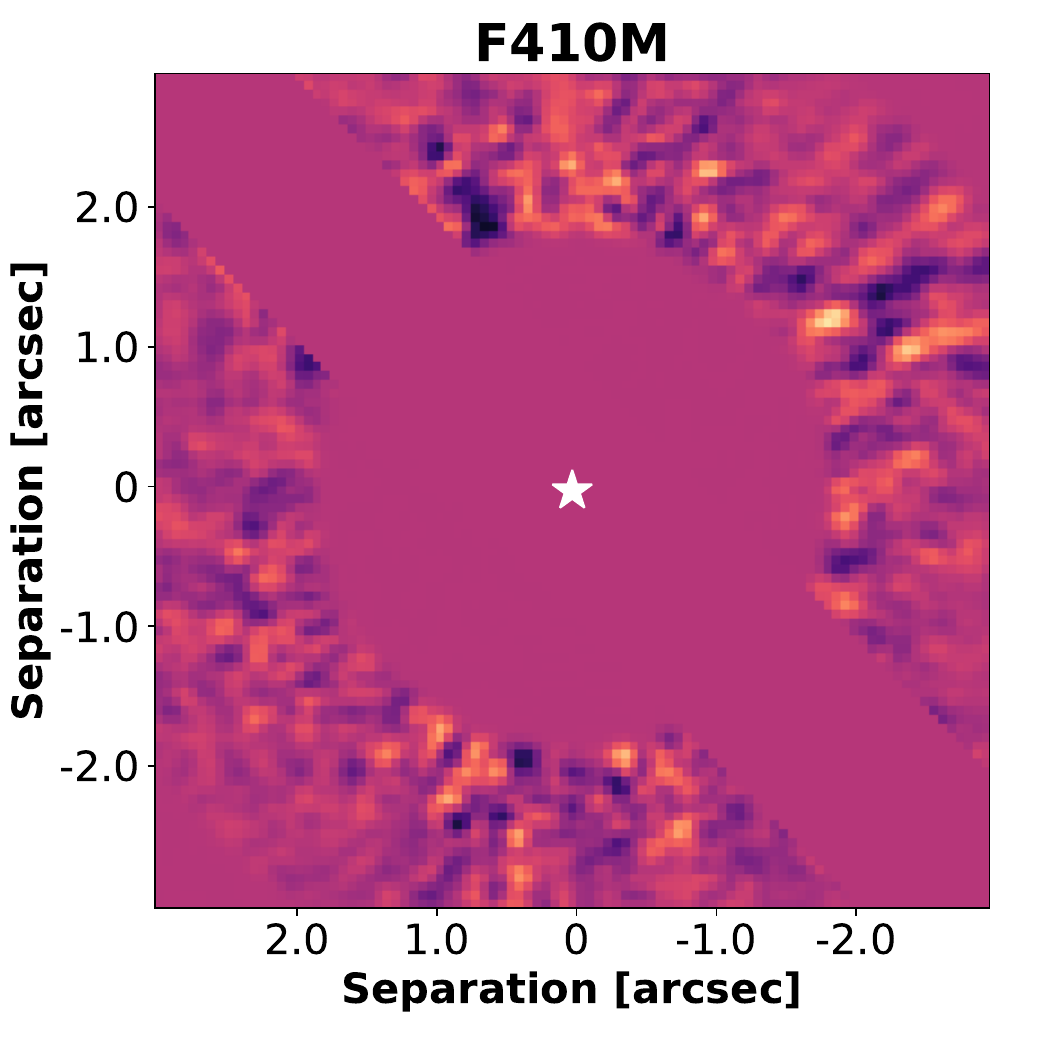}
\caption{Residuals as shown in the left panel of Fig.~\ref{fig:candidate}, with the extended diagonal noise feature masked for clarity. North points to the top, East to the left. 
\label{fig:diag_masked}}
\end{figure}

\section{Injected sources in JWST/NIRCam data}
\label{sec:fake_planet}
\begin{figure}[t!]
\includegraphics[width=1\hsize]{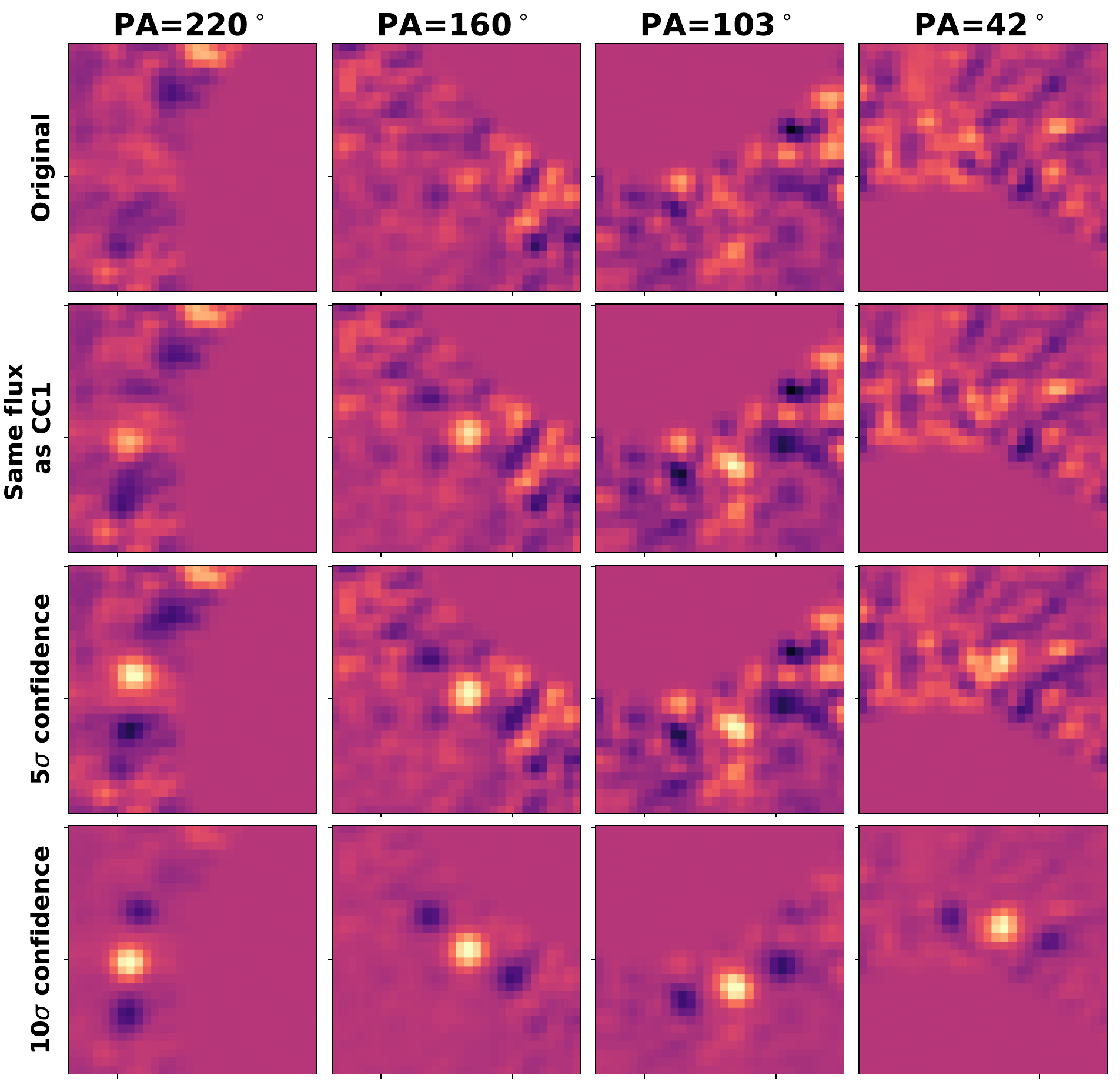}
\caption{{Injected sources in the F410M JWST/NIRCam data at different PAs (reported on top of each column). The top row shows the residuals without injected point source, the second one reports examples of signal with the same flux as CC1, while the third and the fourth rows show higher confidence (5 and 10 $\sigma$) detections. The colorscale is the same for the first three rows, while the limits have been extended for clarity when plotting the fourth row.}
\label{fig:fake}}
\end{figure}

In order to compare the pattern associated with CC1 in the residuals of the F410M NIRCam data, we injected artificial sources at the same separation of CC1 but at different PAs. The zoomed-in regions can be found in Fig.~\ref{fig:fake}, where the top row shows the residuals as reported in Fig.~\ref{fig:candidate} (left panel), the second row shows injected point sources as bright as CC1, while the third and the fourth row show sources at the same location with detection confidences of 5 and 10$\sigma$ respectively. In the first three columns, the point source injected with a similar brightness produces a signal similar to CC1. In the fourth column (PA=$345^\circ$), the signal is very hard to see, as the region is more noisy and stronger residuals than in the rest of the image prevent a clear detection. Due to the more noisy image surrounding, even the 5$\sigma$ signal is hardly recognized in the third row, especially the negative lobes introduced by the two-rolls PSF-subtraction. This indicates that if CC1 were to be located in a different region of the image, in particular along the diagonal noise feature, we would probably have not been able to detect it.

\section{Sensitivity, mass and temperature limits}
Table~\ref{tab:sensitivity} reports the 5$\sigma$ sensitivity limits achieved with JWST/NIRCam in the four filters, together with the best mass limits at each separation (see left panel of Fig.~\ref{fig:mass_teff_limits}) and the lowest $\Teff$ limits at each separation assuming $\Rp=1,2,3~\RJ$ (see right panel of Fig.~\ref{fig:mass_teff_limits}).

\begin{table}[t!]
\centering
\caption{5$\sigma$ Sensitivities, planet mass and effective temperature limits.}
\def\arraystretch{1.1}
\begin{tabular}{lcccccccc}\hline
Sep  & F187N& F200W  & F405N  & F410M  & $\Mp$ & $\Teff~(1~\RJ)$ & $\Teff~(2~\RJ)$ & $\Teff~(3~\RJ)$\\ 
(mas)  & ($\mu$Jy)& ($\mu$Jy) & ($\mu$Jy) & ($\mu$Jy) & ($\MJ$) & (K) & (K) & (K)\\ \hline
0.44 & 5.49 & 114.85 & 1103.8 & 4831.3 & 4.8 & 1115& 928 & 845 \\
0.57 & 2.75 & 60.65  & 642.3  & 2150.7 & 3.8 & 1008& 853 & 782 \\
0.71 & 0.76 & 11.58  & 243.0  & 773.5  & 2.6 & 868 & 750 & 695 \\
0.85 & 0.53 & 5.83   & 95.2   & 127.3  & 2.3 & 829 & 721 & 670 \\
0.98 & 0.39 & 3.31   & 61.1   & 54.1   & 2.1 & 804 & 702 & 654 \\
1.19 & 0.40 & 2.31   & 37.0   & 27.4   & 2.2 & 811 & 707 & 658 \\
1.25 & 0.29 & 1.79   & 32.5   & 19.0   & 2.0 & 783 & 686 & 635 \\
1.40 & 0.30 & 1.54   & 23.3   & 12.4   & 2.0 & 786 & 685 & 595 \\
1.53 & 0.28 & 1.37   & 21.0   & 9.1    & 2.0 & 779 & 649 & 565 \\
1.66 & 0.28 & 0.86   & 14.4   & 8.2    & 2.0 & 778 & 635 & 554 \\
1.80 &      &        & 12.5   & 6.0    & 1.8 & 795 & 606 & 532 \\
1.93 &      &        & 10.6   & 5.3    & 1.6 & 764 & 588 & 518 \\
2.07 &      &        & 10.0   & 3.9    & 1.4 & 723 & 563 & 498 \\
2.21 &      &        & 8.5    & 3.4    & 1.4 & 702 & 550 & 488 \\
2.34 &      &        & 8.5    & 3.4    & 1.3 & 700 & 549 & 487\\
2.48 &      &        & 9.5    & 3.2    & 1.3 & 688 & 542 & 481 \\
2.61 &      &        & 8.4    & 2.5    & 1.2 & 662 & 525 & 468 \\
2.75 &      &        & 6.6    & 1.9    & 1.0 & 635 & 508 & 454 \\
2.89 &      &        & 6.2    & 2.0    & 1.0 & 633 & 507 & 453 \\
3.02 &      &        & 6.7    & 2.0    & 1.0 & 632 & 506 & 453 \\
3.15 &      &        & 7.2    & 1.9    & 1.0 & 627 & 503 & 450 \\
3.29 &      &        & 5.3    & 1.6    & 1.0 & 611 & 493 & 442 \\
3.43 &      &        & 5.2    & 1.3    & 0.9 & 593 & 481 & 433 \\
3.56 &      &        & 4.7    & 1.2    & 0.9 & 585 & 475 & 428 \\
3.70 &      &        & 5.8    & 1.3    & 0.9 & 585 & 475 & 428 \\
3.83 &      &        & 6.3    & 1.4    & 0.9 & 592 & 480 & 432 \\
3.97 &      &        & 4.8    & 0.9    & 0.7 & 554 & 454 & 411 \\ \hline

\end{tabular}\\\vspace{0.2cm}
\label{tab:sensitivity}
\end{table}


\bibliography{sample631}{}
\bibliographystyle{aasjournal}



\end{document}